\magnification=\magstep1
\tolerance=10000

\centerline{{\bf NON-TRIVIAL BEHAVIOUR OF THE SCATTERING}}
\centerline{{\bf AMPLITUDE OF CONTACT-INTERACTING ANYONS}}

\bigskip
\centerline{{\bf Paola Giacconi} $^{a,}$\footnote{$^1$}{E-mail address:
giacconi@bologna.infn.it}, 
{\bf Fabio Maltoni} $^{b,}$\footnote{$^2$}{E-mail address:
fabio.maltoni@cern.ch}, 
{\bf Roberto Soldati} $^{a,}$\footnote{$^3$}{E-mail address:
soldati@bologna.infn.it}}
\medskip
\centerline{$^a$ {\it Dipartimento di Fisica "A. Righi" and Sezione I.N.F.N.,}}
\centerline{{\it via Irnerio 46, 40126 Bologna, Italy}}
\centerline{$^b$ {\it Dipartimento di Fisica and Sezione I.N.F.N.,}} 
\centerline{{\it via F. Buonarroti, 56100 Pisa, Italy}} 
\centerline{{\it and TH Division, CERN, Geneva, Switzerland}}
\bigskip
\noindent
{\bf Abstract}
\medskip
It is shown that the scattering amplitude for contact-interacting
anyons does exhibit a genuine non-perturbative
sector. This means that the corresponding 
perturbative field theoretical formulation,
based on the {\it 2+1} non-relativistic Chern-Simons gauge model coupled to
self-interacting complex scalar field, is not generally able 
to reproduce, order by
order in perturbation theory, the exact result. 
It is proven that the full agreement between the exact 
scattering amplitude and the resummation of the perturbative expansion
of the renormalized 1PI amplitude actually occurs only 
for some continuous sub-family of self-adjoint extensions
of the quantum Hamiltonians, which entail the absence of bound states.
A comparison with previously obtained results is carefully
worked out.

\vfill\eject

\noindent
{\bf I.}\quad
It is well known since twenty years [1] that the Born approximation fails
to reproduce the celebrated Aharonov-Bohm scattering amplitude [2].
Later on, several efforts have been put forward in order to attempt to
reproduce the above amplitude, as well as 
various anyons characteristic features,
by means of perturbative power series expansions [3]. 
There are also recent proposals in the Literature [4-9] to relate,
order by order in perturbation theory, the
strength of the point-like interaction, for two particles on the plane,
with the renormalized coupling of some 
{\it 2+1} dimensional non-relativistic field theoretic models. 
Here we want to show how
that issue appears indeed to exhibit remarkable subtleties; 
moreover we shall be able to point 
out to what extent the above matter holds true and to compare with previous 
results.

A first goal [4] was to put into a close relationship the 
quantum
mechanical scattering amplitude, for two particles  interacting 
via $\delta$-like potential on the plane, 
with the renormalized 1PI four-point function
of a self-interacting non-relativistic complex scalar field model.
Now, it is well known [10] that 
$\delta$-like potentials on the plane are not
mathematically properly defined at the quantum level. 
The correct quantum mechanical formalism to describe point-like interactions 
of two identical point particles with mass $m$
is {\sl only} in 
terms of the self-adjoint extensions of the relative Hamiltonian
$H_0=({\bf p}^2/m)$. This framework will be referred to in the sequel as  
{\sl contact-interaction}.

This leads [10] to the non-trivial scattering amplitude - which is entirely due 
to the $S$-partial wave - given by
$$
\eqalign{
\sqrt{\pi ip}f (p,E_0) & \equiv e^{2i\delta_0}-1
=\exp\left\{2i\arctan{\pi\over 
\ln (-p^2/mE_0)}\right\}- 1\cr
& = {2\pi i\over \ln (-p^2/mE_0) - i\pi}\ ;\cr}
\eqno(1)
$$
here $p$ is the modulus of the relative momentum, while $E_0=E_B< 0$ is the 
bound state energy and/or the opposite of the resonance's energy  
$E_{{\rm res}}=-E_0$, which labels the one-parameter continuous family of 
the self-adjoint extensions of $H_0$ disclosing, thereby, the presence 
of some non-trivial 
attractive interaction due to the boundary condition at the origin.
As a matter of fact, it turns out that, for any given non-trivial behaviour 
at the origin of
the square integrable wave function, a bound state 
{\sl always} exists [10] which specifies,
in the most natural and physical manner, the corresponding self-adjoint
extension of $H_0$. 
Notice that the limiting case $E_B\to -\infty$ -
the so called Friedrichs' extension corresponding to regular wave functions -  
leads to a vanishing amplitude, {\it i.e.} to the switching off of 
the contact-interaction (free particle). Consequently, we can conclude that
in two spatial dimensions the contact-interaction, which leads to
non-trivial physical effects, turns out always to be of attractive nature.

Now, one can try rather easily to identify the quantum mechanical
momentum scale $\sqrt{m|E_B|}$, due to the presence of contact-interaction,
with the subtraction point $\mu$ 
of a renormalizable field theoretic model. The question is, therefore, to 
carefully verify whether such an identification is actually there, taking into
account the close relationship between the quantum mechanical 
scattering amplitude and the renormalized
1PI four-point function - in the center of mass frame -
of a non-relativistic field theoretic model: namely,
$$
\sqrt{\pi ip}f(\varphi,p)={m\over 4}\Gamma^{(4)}_{\rm R}(p,\varphi)\ ,
\eqno(2)
$$
which holds true in {\it 2+1} dimensions.

The model is that of a self-interacting complex scalar field, as  
described by the following renormalized lagrangean
density in {\it 2}$\omega${\it +1} dimensions
$$
{\cal L}_R  \equiv {\cal L}\ +\ {\cal L}_{{\rm c.t.}}
 = \phi^*\left(i\partial_t + {1\over 2m} \triangle\right)\phi
- \mu^{2\epsilon}{\lambda_0\over 4m}(\phi^*\phi)^2\ ,
\eqno(3)
$$
where $\triangle\equiv \partial^2$, the bare coupling $\lambda_0$ being
determined, order by order in perturbation theory, by
$$
\eqalign{
\lambda_0 & \equiv
a_0(\lambda)+\sum_{k=1}^\infty a_k(\lambda){1\over \epsilon^k}\cr
          & = \lambda+F(\epsilon,m/\mu)
+{\lambda^2\over 8\pi
\epsilon} + {\lambda^3\over 64\pi^2\epsilon^2}+
{\cal O}\left({\lambda^4\over \epsilon^3}\right)\ ,\cr}
\eqno(4)
$$
$\epsilon\equiv \omega -1$ and $\mu$ is the conventional mass
parameter within dimensional regularization. 
Here, the quantity $F(\epsilon,m/\mu)$ is the
usual arbitrary finite part of the counterterm
(analytic for $\epsilon\to 0$), which has to be 
fixed by some renormalization prescription, while all the coefficients of
the higher poles $(1/\epsilon^k)$ can be readily computed from renormalization 
group equations.

Now, it is an easy exercise to show that, if the finite part of the 
counterterm for the one-loop
1PI four-point function - which turns  out to be the only renormalization
part - is chosen to be equal to zero (MS-scheme),
then the exact renormalized 1PI four-point function, 
in two spatial dimensions    
and in the center of mass frame, reads
$$
\Gamma^{(4)}_{{\rm R}}(p/\mu^\prime)=
\left(-i{\lambda\over m}\right)
{1\over 1-(\lambda/4\pi)\left\{\ln (p/\mu^\prime)-
i(\pi/2)\right\}}
\ ,
\eqno(5)
$$
where $\mu^\prime$ is
defined from $\ln\mu^\prime =\ln\mu+(1/2)[\ln (4\pi)-\gamma_E]$,
$\gamma_E$ being the Euler-Mascheroni constant.
According to Ref. [4], a comparison between Eq.s~(1) and (5) 
allows us to recognize the following correspondences, 
taking Eq.~(2) into account: namely,
$$
{8\over m}\tan\delta_0 (p,E_B)={(8\pi/m)\over \ln(-p^2/mE_B)}
=i\Gamma^{(4)}_{{\rm R}}(ip/\mu^\prime)\ , 
\eqno(6)
$$
$$
{4\pi\over \lambda}=\ln\left({\sqrt{m|E_B|}\over \mu^\prime}\right)\ ;
\eqno(7)
$$
here $\delta_0(p,E_B)$ denotes the phase shift associated to the scattering
amplitude of Eq.~(1). 
We notice that, owing to the fact that only attractive 
pure contact-interactions on the plane are non-trivial, $i.e.$ a bound
state is always present [4], [10], there is no relationship between the
nature of the two-dimensional pure contact-interaction and 
the sign of the scalar
self-interaction, at variance with one- and three-dimensional cases [11].
In other words, the sign of the renormalized coupling $\lambda (\mu^\prime)$
is merely determined by the relative magnitude between the physical scale
$\sqrt{-mE_B}$ and the definite (but arbitrary) choice of the subtraction
scale $\mu$, the attractive nature of the contact-interaction being
anyway understood.   

It is apparent that Eq.~(7) precisely provides the above mentioned
relationship among the renormalized coupling $\lambda$, within
the MS-renormalization prescription, the mass scale $\mu$
and the bound state energy $E_B$. 
It is worthwhile to notice that,
from Eq.~(7) and taking into account that the bound state energy
$E_B$ is a {\sl physical} parameter - {\it i.e.} 
$\mu$-independent - one immediately
gets [4] that the $\beta$-function is exactly given by 
$$
\beta (\lambda)={\lambda^2\over 4\pi}\ .
\eqno(9)
$$
This exact value can also be obtained from perturbation theory, to any order,
by noticing that there are no higher order corrections - in the running coupling
$\lambda (\mu^\prime)$ - to the simple pole $(1/\epsilon)$ in Eq.~(4). 
Furthermore, it appears that Eq.~(5) truly corresponds to the sum of all 
the renormalized
Feynman graphs and turns out to be, as it does, {\sl analytic} function of the
renormalized coupling parameter $\lambda$.

The actual correspondences based upon Eq.s~(2) or (6),
appear to be firmly established as they stand;
it is natural to see, therefore, whether there exists 
a generalization in the presence of the Aharonov-Bohm interaction, {\it i.e.}
whether the scattering amplitude for 
non-relativistic contact-interacting (or colliding)
anyons may be exactly reproduced, after resummation of the perturbative 
series of some suitable field theoretic model [6-9].
\bigskip
\noindent
{\bf II.}\quad
Let us in fact consider the quantum mechanical scattering amplitude, when
the Aharonov-Bohm (AB) gauge potential $A_j (x_1,x_2)=\alpha\varepsilon_{jk}
(x_k/r^2),\ j,k=1,2,\ \varepsilon_{12}=1,\ r^2=x_1^2+x_2^2$, is switched on.
We can restrict ourselves to the interval $-1<\alpha <0$ cause, as it is
well known, for integer values of $\alpha$ the Aharonov-Bohm quantum
Hamiltonian is gauge equivalent to the previously discussed
pure contact-interaction Hamiltonian.
There, once again,
the contact-interaction is described by the self-adjoint extensions
\footnote{$^\sharp$}{As the AB hamiltonian operator
 has deficiency indices (2,2), 
the most general form of the self-adjoint extensions
of the AB hamiltonian operator is provided by a four parameter family, whose 
elements do not commute, in general, with the angular momentum operator [12].}
of the relative hamiltonian symmetric operator $H_\alpha=
(1/m)[{\bf p}-{\bf A}({\bf r})]^2$.

As we are here interested in the boson scalar like matter, we have to
consider the radial differential operator with vanishing angular
momentum: namely,
$$
h_s(\alpha)={\hbar^2\over m}\left\{-{d^2\over dr^2} -{1\over r}
{d\over dr}+{\alpha^2\over r^2}\right\}\ .
\eqno(10)
$$
The general solution of the eigenvalue equation for the $S$-wave can
be written in the form
$$
\psi_s(pr)=AJ_{|\alpha|}(pr)+BN_{|\alpha|}(pr)\ ,
\eqno(11)
$$
with $p=\sqrt{mE}$ labelling the continuous part of the spectrum, 
$J_{|\alpha|}(pr)$ and $N_{|\alpha|}(pr)$ being
the Bessel and Neumann functions respectively. Notice that the coefficients
$A$ and $B$ can always be chosen to be real without loss of generality.

It turns out [10],[13],[14] that the one parameter 
continuous family of the
self-adjoint extensions of the radial Hamiltonian of zero angular
momentum can be described in terms of the quantity $E_0$ according to
$$
\gamma\equiv {A\over B}\sin\pi\alpha - \cos\pi\alpha= 
{\rm sgn}(E_0)\left({p^2\over m|E_0|}\right)^\alpha\ ,
\eqno(12)
$$
where $-\infty\le E_0< +\infty$, the energy scale $E_0$ being here evidently
assumed to be independent from the magnetic flux and/or statistical
parameter $\alpha$, as it has to label the self-adjoint extensions.
In all and only the cases 
in which $-\infty < E_0 < 0$, there always exists a 
bound state whose energy is precisely $E_B=E_0$. 
Furthermore, the case $E_0= -\infty$ corresponds to the original 
Aharonov-Bohm quantum Hamiltonian [2].

It also happens that
the analysis of the stationary scattering states leads to the 
following $S$-wave phase shift
$$
\tan\left\{\delta_0(p,\alpha,E_0)-{\pi\over 2}\alpha\right\}
\equiv -{B\over A}= \sin\pi|\alpha|
\left\{{\rm sgn}(E_0)\left({p^2\over m|E_0|}\right)^\alpha
+\cos\pi|\alpha|\right\}^{-1}\ .
\eqno(13)
$$
Then, an easy exercise leads to the scattering partial $S$-wave amplitude
$$
\sqrt{\pi ip}f_0(p,E_0;\alpha)=\left(1-e^{i\pi\alpha}\right)
{1-{\rm sgn}(E_0)(p^2/m|E_0|)^\alpha\over
\exp\{i\pi\alpha\}+{\rm sgn}(E_0)(p^2/m|E_0|)^\alpha}\ .
\eqno(14)
$$
It is immediate to check that, in the limit $\alpha\to 0$, the 
pure contact-interaction amplitude of Eq.~(1) is readily recovered
from  Eq.~(14) when $-\infty\le E_0<0$, whereas the expression (14) 
indeed vanishes in the above limit when $0\le E_0< +\infty$.
\footnote{*}{We observe that
the value $E_0=0$ corresponds to a well defined self-adjoint 
Hamiltonian, whose domain is that one of pure irregular wave functions at the
origin.}

The open question is now to establish whether the above quantum mechanical
amplitudes can be exactly reobtained order by order in perturbation theory, 
taking the general relation of
Eq.~(2) into account, starting from the field theoretic model described by the 
renormalized lagrangean density in {\it 2$\omega$+1} dimensions
$$
{\cal L}_R ={\cal L}_{{\rm matter}} + {\cal L}_{{\rm CS}}\ ,
\eqno(15)
$$
corresponding to a non-relativistic charged scalar field interacting with a
Chern-Simons gauge field [5]. Here we actually have
$$
\eqalignno{
&{\cal L}_{{\rm matter}}=\phi^*\left(i\partial_t+eA_0\right)\phi
+ {1\over 2m}\left|i\nabla\phi+e{\bf A}\phi\right|^2 
- \mu^{2\epsilon}{\lambda_0\over 
4m}(\phi^*\phi)^2\ , &(16.a)\cr
&{\cal L}_{{\rm CS}}={\kappa\over 2}\varepsilon_{jl}A_l
\left(\partial_tA_j-\partial_jA_0\right)\ , &(16.b)\cr}
$$ 
with
$$
\lambda_0\equiv 
a_0(\lambda,\nu)+\sum_{k=1}^\infty a_k(\lambda,\nu){1\over
\epsilon^k}\ ,\quad \nu\equiv {e^2\over \kappa}=2\pi\alpha\ .
\eqno(17)
$$
To be quite general, we understand that the scalar self-interaction
renormalized coupling $\lambda$ and the CS coupling $\nu$,
which does not renormalize, are independent free parameters.
In so doing, the renormalized lagrangean density (3) is
recovered in the limit $\nu\to 0$. 

It is possible to show,
from Feynman's rules and the power counting criterion, 
that (15) is stable under 
radiative corrections, $i.e.$ there is no need of any further counterterm, but 
scalar self-interaction, to make Green's functions finite to all orders in
perturbation theory [15].
It is important to stress that, within the present model, 
the scalar self-interacting bare 
coupling is necessarily generated by the
radiative corrections induced from the minimal coupling with the Chern-Simons
gauge potential. 

Now, starting from the lagrangean (15),
the perturbative evaluation of the 
partial $S$-wave renormalized scattering amplitude
 leads to, up to the 
two-loop approximation [5-6] and in the MS-renormalization-scheme,
$$
\eqalign{
\left.\Gamma^{(4)}_{{\rm R},s}(p/\mu^\prime,\nu)\right|_{2-loop}
 &=-i{\lambda\over m}-{\nu^2\over 2m}-
{i\over m}{\lambda^2-4\nu^2\over 4\pi}
\left(\ln{p\over \mu^\prime}-i{\pi\over 2}\right)\cr
&-{i\lambda\over 4\pi m}{\lambda^2-4\nu^2\over 4\pi}
\left(\ln {p\over \mu^\prime}-i{\pi\over 2}\right)^2
+i{\lambda\nu^2\over 24m}\ ,\cr}
\eqno(18)
$$
together with the relationship
$$
\lambda_0(\epsilon)=
\lambda+{\lambda^2-4\nu^2\over 8\pi}{1\over \epsilon}
+{\lambda\over 8\pi}{\lambda^2-4\nu^2\over 8\pi}{1\over \epsilon^2}
+{\cal O}\left({1\over \epsilon^3}\right)\ .
\eqno(19)
$$

First we notice that, once again, owing to the absence of higher order 
corrections to the coefficient of the simple pole in $(1/\epsilon)$,
the $\beta$-function is exactly provided by the expression
$$
\beta (\lambda)={\lambda^2-4\nu^2\over 4\pi}\ ,
\eqno(20)
$$
which means that perturbative scale invariance occurs at the critical values
$\lambda^{(\pm)}_{{\rm cr}}=\pm 2\nu=\pm 4\pi\alpha$, according to Ref. [5]. 
It should be stressed that those
critical values are obtained within the framework of field theoretic
perturbation theory; on the other hand, we recall that 
the exact quantum mechanical amplitude of Eq.~(14)
exhibits non-perturbative 
scale invariance if and only if $E_0= -\infty,\ E_0=0$.

Now the key point.
A straightforward comparison between the quantum mechanical 
amplitude of Eq.~(14) and the field theoretical perturbative expansion 
of Eq.s~(18-19), allows us to establish the domain in the parameter 
space $(\alpha , \lambda)$ in which the two formulations appear to
be exactly equivalent. This analysis shows that the exact correspondence
takes place only in the following two cases:
\medskip

$a)$ $\alpha = 0$,\ $\forall\lambda\in{\bf R}$\ (pure contact-interaction);

$b)$ $\lambda =0$,\ $\forall\alpha\in ]-1,0]$\ (CS minimal coupling),
     {\bf provided} $0<E_0<+\infty$.
\medskip
\noindent
Let us in fact consider the limit $\alpha\to 0$ of Eq.~(18); then the 
expansion in powers of $\lambda$ of the pure contact-interaction 
amplitude of Eq.~(5) is correctly recovered.
If, instead, we let $\lambda \to 0$, then the expansion in powers
of $\alpha$ of Eq.~(14) is indeed obtained only in the domain 
$0<E_0<+\infty$,
which is characterized by the absence of bound states.
This means that,
strictly speaking, only the amplitudes corresponding to the 
sub-family of self-adjoint Hamiltonians with purely continuous spectrum
and non-trivial scaling behavior are actually reproduced 
by the renormalized minimally coupled Chern-Simons perturbative field theory
in the MS-scheme, $i.e.$ with $a_0(\lambda,\nu)=0$ and $a_k = a_k(\nu)$,
provided we identify $\mu^\prime=\sqrt{m E_0}$.

On the contrary, 
the expansion in powers of $\alpha$ - the strength of
the AB interaction - of the exact quantum mechanical amplitude (14) in the
attractive domain $E_B=E_0<0$, where a bound state is always present, reads
$$
\eqalign{
& \sqrt{\pi ip}f_0 (p,E_B;\alpha)=\cr
& {(im/4)T(p,E_B)\over 1-(im/8)T(p,E_B)}+{4i\pi^2\alpha^2\over
3mT(p,E_B)}{1+(im/8)T(p,E_B)\over 1-(im/8)T(p,E_B)}+{\cal O}(\alpha^4)\ ,\cr}
\eqno(21)
$$
where
$$
T(p,E_B)\equiv {(8\pi/m)\over \ln(-p^2/mE_B)}\ .
$$
Now, since the first term in the RHS of the expansion is just
$\sqrt{\pi ip}f_0 (p,E_B;\alpha=0)$, {\it i. e.} the amplitude of Eq.~(1),
it follows from Eq.~(5-7) that the quantity $T(p,E_B)$ must be given by 
$$
T(p,E_B)={(\lambda/m)\over 1-(\lambda/4\pi)\ln(p/\mu^\prime)}\ ,\quad
E_B=-{\mu^{\prime 2}\over m}\exp\left\{{8\pi\over \lambda}\right\}\ .
\eqno(22)
$$

As a consequence, it is absolutely manifest that 
the coefficient of $\alpha^2$ in Eq.~(21) is
{\sl not analytic} in the renormalized coupling $\lambda$. 
Thereby, the resummation of
the perturbative expansion of $\Gamma^{(4)}_{{\rm R},s}(p/\mu^\prime,\alpha)$,
whose two-loop value is provided by Eq.~(18),
will never be such to fully reproduce the exact formula (14), 
which turns out to contain
truly non-perturbative effects (the AB effect in the presence of the 
attractive contact-interaction).

We stress once again that the latter conclusions lies on the assumed 
independence from $\alpha$ of the energy scale $E_0$, which labels the 
one-parameter family of self-adjoint Hamiltonians. On the other hand, this
very same assumption is mandatory in order to correctly reproduce, in the
limit $\alpha\to 0$,  the pure contact-interaction amplitude of Eq.~(1).
An alternative possibility has been considered in 
Ref.s~[6], [7], [8], where it is instead tacitly 
assumed that the contact-interaction
disappears after the switching off of the AB interaction,
as we shall further discuss in the sequel.

To sum up, in the parameter plane $(\alpha ,\lambda)$ the precise
correspondence between perturbative quantum field theory and 
non-perturbative quantum mechanics
is rigorously established only on the coordinate's axes  
$\alpha=0$, $\lambda=0$ and only for the continuous sub-family of 
self-adjoint extensions which do not admit bound states, namely
$0<E_0<+\infty$. 

\bigskip
\noindent
{\bf III.}\ It is very instructive to compare our result with the 
treatments given in the recent Literature [5-9] and to carefully
discuss why different points of view and conclusions do indeed arise. 

In order to do this,
let us first briefly recollect and comment some of the most frequently
used parametrizations and conventions concerning the
self-adjoint extensions of the Aharonov-Bohm relative 
radial hamiltonian operator of vanishing angular momentum (see Eq.~(10)).
Those extensions have to be encoded into the quantity
 (see Eq.s~(11) and (12))
$$\gamma\equiv {A\over B}\sin\pi\alpha - \cos\pi\alpha\ .
\eqno(23)
$$
The condition for the absence or presence of a bound state,
$\forall \alpha\in ]-1,0]$, reads respectively:
$\gamma =-\infty ,\ \gamma \ge 0$ (absence) and 
$-\infty < \gamma < 0$ (presence).

As the phase shift of the $S$-partial wave is provided by
$$
\tan\left\{\delta_0+{\pi\over 2}|\alpha|\right\}=-{B\over A}=
{\sin\pi|\alpha|\over \cos\pi|\alpha|+\gamma}\ ,
\eqno(24)
$$
the $S$-partial wave 
scattering amplitude can be immediately written in the form
$$
\sqrt{\pi ip}f_0(\alpha ; [\gamma])=\left(1-
e^{-i\pi|\alpha|}\right)
{1-\gamma\over e^{-i\pi|\alpha|}+\gamma}\ .
\eqno(25)
$$
Notice that the limit $\gamma\to -\infty$ 
corresponds to the original AB amplitude [2].

In order to exhibit some explicit formula for the scattering amplitude, it is
mandatory to specify the quantity $\gamma$ as a function of the
following {\sl independent variables}: the modulus $p$ of
the center of mass momentum, the magnetic flux (or statistical)
parameter $\alpha$ and some further parameter, which labels the continuous
family of self-adjoint extensions, which necessarily involves the presence
of an additional momentum scale (or length scale).
To this aim, two alternative basic choices are available in the Literature: 
let us briefly analyse both of them.

The first possibility is the one we have previously introduced
(see Eq.~(12)), which is equivalent to the one of 
Ref.~[14]: namely,
$$
\gamma (\alpha, E_0) = {\rm sgn}(E_0)
\left({\sqrt{m|E_0|}\over p}\right)^{2|\alpha|}\ ,
\eqno(26)
$$
with $-\infty\le E_0< +\infty$ and $E_0$ is understood to be 
$\alpha$-independent. We notice that such an energy scale is a physical
observable quantity: it represents the resonance energy (when $E_0 >0$ and
$-1<\alpha <-{1\over 2}$), or the slope of the cross section (when
$E_0>0$ and $-{1\over 2}<\alpha <0$), or the bound state energy (when
$E_0<0$).

The second possibility is the parametrization used in Ref.~[6]: namely,
$$
\gamma (\alpha,w) = {1\over w}
\left({2\hbar\over pR}\right)^{2|\alpha|}
{\Gamma (1+|\alpha|)\over \Gamma (1-|\alpha|)}\ ,
\eqno(27)
$$
with $R$ is some fixed length (say 1 cm) and $w\in {\bf R}$ 
independent from $|\alpha|$. Here, the dimensionless quantity $w$ is 
related to the behavior of the wave function at the origin and, thereby,
it does not represent a directly observable quantity.

Now, it is absolutely crucial to gather the following subtle point.
For a given fixed non-zero value of the magnetic flux (or statistical parameter)
$\alpha$, Eq.s~(26) and (27) give perfectly equivalent labelling of the 
continuous one-parameter family of self-adjoint extensions of the $S$-wave
radial hamiltonian, according to Von Neumann's theorem. 

But, if we want to establish some correspondence with perturbative quantum
field theory, we need more, as we must look at the scattering amplitude as  
analytic function of the variable $\alpha\in ]-1,0]$.
In so doing, it will be clear, as we shall explain below, that the two options
of Eq.s~(26) and (27) actually describe physically different situations and,
not surprisingly, they lead to quite different conclusions concerning the 
possibility to fully reproduce the scattering amplitude within perturbative
quantum field theory.

{\it i)}\  
If we restrict ourselves to the continuous sub-family of self-adjoint
extensions labelled by $0<\gamma < +\infty$, it is easy to realize that
both parametrizations lead to equivalent descriptions in the following sense. 
The pure CS model $\lambda=0,\
\mu^\prime =\sqrt{mE_0}$ or such an $\alpha$-dependent renormalized 
scalar coupling 
$$
\lambda (\alpha)=4\pi|\alpha|{1-w\over 1+w}\ ,\quad \mu^\prime=
\sqrt{4\pi}(2\hbar/R)\exp\{-(3/2)\gamma_E\}\ ,
\eqno(28)
$$ 
as proposed in Ref.~[6], both reproduce,
order-by-order in the renormalized couplings, the very same amplitude (25),
when $0<\gamma < +\infty$ that means $w>0$,
up to some one-to-one redefinition of the parameters.
Notice that, if $0<\gamma < +\infty$, 
the amplitude vanishes when $\alpha\to 0$,
{\it i.e.} when the CS-AB interaction is turned off.

{\it ii)}\
On the contrary, within the complementary continuous sub-family of
self-adjoint extensions labelled by $-\infty\le\gamma\le 0$, that
includes the standard AB case,
the two parametrizations
describe truly different physical situations. As a matter of fact, according 
to our choice (26), the bound state energy $E_0=E_B<0$ - a physical
observable quantity - is assumed to be independent from the magnetic flux
$\alpha$ and, thereby, it becomes the natural parameter which labels the
self-adjoint extensions.

Owing to this choice (26), on the one hand
the amplitude (21) is not analytic function of the
renormalized scalar self-interaction renormalized coupling $\lambda$ but,
on the other hand, from the very same choice (26), one
does correctly reproduce the non-vanishing pure 
contact-interaction case of Eq.~(1), when $\alpha\to 0$.

We stress once again that, in so doing, the underlying CS gauge field
theory involves a renormalized scalar self-interaction 
$\lambda$ truly independent from 
the scalar-CS minimal coupling $\nu$ - which does not renormalize -
as it must be on general ground:
namely,
$\lambda(\alpha=0)\not=0$ at variance with Eq.~(28). 

Alternatively, 
if the amplitude (25) is expressed by means of Eq.~(27) with
$-\infty<w< 0$, then the bound state energy reads:
$$
E_B(\alpha ,w) =-\left(-{1\over w}\right)^{1/|\alpha|}
\left({4\hbar^2\over mR^2}\right)\left[{\Gamma(1+|\alpha|)\over
\Gamma(1-|\alpha|)}\right]^{1/|\alpha|}\ .
\eqno(29)
$$
Now, it is quite clear that, under the assumption that $w$ is 
$\alpha$-independent - what is explicitly done in Ref.s~[6] and [9] -
the bound state energy, which is a physical quantity,
becomes $\alpha$-dependent, at variance with what happens in the previous
approach: this is the reason why it can no longer be used, in this approach,
 as an independent
parameter to label the continuous family of self-adjoint 
extensions. Furthermore, in the limit $\alpha\to 0$ the bound state 
energy (29) goes to $-\infty$, which means that the pure contact-interaction
case can no longer be recovered. 
Nevertheless it is remarkable that, thanks to Eq.~(28), the sub-family of 
the scattering amplitudes which entail the presence of a bound state
({\it i.e.} $-\infty<w<0$) might be reobtained after resummation of the
perturbative field-theoretic expansion, but for the singular case $w=-1$.
It would be very interesting to verify whether the latter circumstance 
still occurs in the $N$-particle sectors (with $N>2$) [16], in the presence
of additional interactions other than the AB potential and in the 
relativistic case [17].

To sum up, we can draw the following conclusions:

{\it i)}\ in the absence of bound states, the exact solution of the quantum
mechanical two-body problem can always be reobtained after resummation of
the perturbative expansion from the Chern-Simons field theoretic model in
the non-relativistic case.

{\it ii)}\ in the presence of bound states, two alternative 
situations do occur:
\par\noindent
{\tt a)}\ if the physical bound state energy $E_B$ results to be 
{\sl independent} from the magnetic flux (or statistical) parameter $\alpha$,
then the renormalized coupling $\lambda$ is also $\alpha$-independent and
the scattering amplitude is not analytic in $\lambda$, albeit
the pure contact-interacting case is correctly recovered in the limit 
$\alpha\to 0$;
\par\noindent
{\tt b)}\ if the physical bound state energy $E_B$ results to be some suitable
function of $\alpha$, then the renormalized coupling $\lambda$ is also 
$\alpha$-dependent and
the exact quantum mechanical scattering amplitude
can be reobtained after resummation from the CS field theoretic 
perturbative approach, but for the special case 
$E_B(\alpha , w=-1)=
-\left({4\hbar^2\over mR^2}\right)\left[{\Gamma(1+|\alpha|)\over
\Gamma(1-|\alpha|)}\right]^{1/|\alpha|}$; 
however, the pure contact-interacting
case cannot be recovered in the limit $\alpha\to 0$.

As a final comment, we can say that, since the bound state energy $E_B$ and
the magnetic flux $\alpha$ are observable quantities, their actual relationship
can be in principle experimentally checked. Therefore, the above discussed 
alternative possibilities of describing the model in terms of perturbative 
quantum field theory do eventually concern the concrete physical 
framework to which the model itself could actually be applied.
\bigskip
\noindent
{\bf Acknowledgments}
\medskip
This work is partially supported by a grant MURST - quota 40\%. We thank
the CERN Theory Division for the kind hospitality during the last stage of
the work. We wish to thank R. Adami for helpful
discussions and D. Bak for useful correspondence. One of the authors (R. S.)
received great benefits from conversations with M. Consoli, C. Manuel,
S. Korenblit, V. Leviant, D. Naumov, S. Ouvry and R. Tarrach.
We are specially indebited to G. Amelino-Camelia for his stimulating 
contribution to improve a previous version of the present paper.

\bigskip
\noindent

{\bf References}

\medskip

\item{[1]}\ E. Corinaldesi, F. Rafeli : Am. J. Phys. {\bf 46} (1978) 1185.

\item{[2]}\ Y. Aharonov, D. Bohm : Phys. Rev. {\bf 115} (1959) 485.

\item{[3]}\ C. Chou : Phys. Rev. {\bf D 44} (1991) 2533;\par
            G. Amelino-Camelia : Phys. Lett. {\bf 286B} (1992) 97;\par
            S. Ouvry : Phys. Rev. {\bf D51} (1994) 5296;\par
            C. Manuel, R. Tarrach : Phys. Lett. {\bf 328B} (1994) 113;\par
            G. Amelino-Camelia : Phys. Rev. {\bf D51} (1995) 2000.

\item{[4]}\ O. Bergmann : Phys. Rev. {\bf D46} (1992) 5474.

\item{[5]}\ O. Bergmann, G. Lozano : Ann. Phys. (N.Y.) {\bf 229} (1994) 416.

\item{[6]}\ G. Amelino-Camelia, D. Bak : Phys. Lett. {\bf 343B} (1995) 231.

\item{[7]}\ S.-J. Kim : Phys. Lett. {\bf 343B} (1995) 244.

\item{[8]}\ D. Bak, O. Bergmann : Phys. Rev. {\bf D51} (1995) 1994.

\item{[9]}\ S.-J. Kim, C. Lee : Phys. Rev. {\bf D55} (1997) 2227.

\item{[10]}\ S. Albeverio, F. Gesztesy, R. Hoegh-Krohn, H. Holden :
            ``{\it Solvable Models in Quantum Mechanics}'' (Springer-Verlag,
            New York, 1988).

\item{[11]}\ R. Jackiw, in {\it M.A.B. Beg Memorial Volume}, A. Ali and
            P. Hoodbhoy Eds. (World Scientific, Singapore, 1991).

\item{[12]}\ R. Adami, A. Teta : preprint 97/4, Dip.to di Matematica, Roma
            {\sl "La Sapienza"} (1997).

\item{[13]}\ P. Giacconi, F. Maltoni, R. Soldati : Phys. Rev. {\bf D53}
            (1996) 952.

\item{[14]}\ C. Manuel, R. Tarrach : Phys. Lett. {\bf 268B} (1991) 222.

\item{[15]}\ D. Z. Freedman, G. Lozano, N. Rius : Phys. Rev. {\bf D49} (1994)
             1054.

\item{[16]}\ A. N. Vall, S. E. Korenblit, V. M. Leviant, D. V. Naumov,
             A. V. Sinitskaya : {\tt hep-th/9710108}.
\item{[17]}\ K. Chadan, N. N. Khuri, A. Martin, T. T. Wu :
             {\tt hep-th/9805036}.

\vfill\eject\end